\documentclass[useAMS,usenatbib]{mnras}

\usepackage{graphicx}

\def\aap{A\&A}

\def\apjs{ApJS}

\def\lap{\hbox{\hspace{4.3mm}}
         \raise1.5pt \vbox{\moveleft9pt\hbox{$<$}}
         \lower2.5pt \vbox{\moveleft9pt\hbox{$\sim$ }}
         \hbox{\hskip 0.02mm}}
\def\gap{\hbox{\hspace{4.3mm}}
         \raise1.5pt \vbox{\moveleft9pt\hbox{$>$}}
         \lower2.5pt \vbox{\moveleft9pt\hbox{$\sim$ }}
         \hbox{\hskip 0.02mm}}

\title[Nonbirefringent polarization modes in pulsars]{Orthogonal
polarization modes of nonbirefringent origin in a geometric radio pulsar signal model}

\author[J.~Dyks, L.~Saha, A.~Frankowski]%, et al.]%, M.~Serylak, S.~Os{\l}owski, et al.]
{J.~Dyks$^1$, L.~Saha$^2$ and A.~Frankowski$^1$  
\\
$^1$Nicolaus Copernicus Astronomical Center, Polish Academy of Sciences, Rabia\'nska 8, 87-100, Toru\'n,
Poland\\
$^2$Center for Astrophysics $|$ Harvard \& Smithsonian, Cambridge, MA 02138, USA. 
}
\begin{document}

%\date{Accepted .... Received 2015 August 10; in original form 2015 July 23}
%\date{Accepted . Received ...; in original form 2020 Sep 22}

\date{Accepted 2026 March 12. Received 2025 October 8; in original form 2025
October 8}
%\pagerange{\pageref{firstpage}--\pageref{lastpage}} \pubyear{2002}

\maketitle

\label{firstpage}

\begin{abstract} 
Motion of plasma in pulsar magnetospheres essentially follows bent magnetic field
lines which may result in radiation beams sweeping past the observer's sightline. 
We show that under certain conditions such a passing beamlet can produce two orthogonal
polarization modes (OPMs) solely from averaging of radiative contributions, without involving birefringence. 
This requires that radiation from different stages of the passage is
contributed incoherently, hence that the timescale for the coherence is
much shorter than the typical sampling time. The model produces two OPMs of
comparable amount, with a wide range of (single-pulse) polarized fractions, which is
consistent with general observed pulsar properties. Under the specific
conditions required (essentially incoherent summation of signal at appropriate time scale) 
the averaging effects alone can thus lead to the appearance of two orthogonal
polarization tracks with OPM jumps. 
%{\bf The observed OPMs result from symmetry of emitters, in
%particular of elnogated elementary plasma streams.} 
% Since the beamlet passage is mathematically equivalent to integration along an elongated stream, 
%The observed OPMs can be considered a direct consequence of the 
%stream-shaped (fan-beam) geometry of elementary magnetospheric radio emitters.} 
\end{abstract}


\def\lap{\hbox{\hspace{4.3mm}}
         \raise1.5pt \vbox{\moveleft9pt\hbox{$<$}}
         \lower1.5pt \vbox{\moveleft9pt\hbox{$\sim$ }}
         \hbox{\hskip 0.02mm}}

\def\rwobs{R_W}
\def\rwcon{R_W}
\def\rwstr{R_W}
\def\winobs{W_{\rm in}}
\def\woutobs{W_{\rm out}}
\def\phm{\phi_m}
\def\phmi{\phi_{m, i}}
\def\thm{\theta_m}
\def\dres{\Delta\phi_{\rm res}}
\def\win{W_{\rm in}}
\def\wout{W_{\rm out}}
\def\rin{\rho_{\rm in}}
\def\rout{\rho_{\rm out}}
\def\phin{\phi_{\rm in}}
\def\phout{\phi_{\rm out}}
\def\xin{x_{\rm in}}
\def\xout{x_{\rm out}}

\def\thmin{\theta_{\rm min}^{\thinspace m}}
\def\thmax{\theta_{\rm max}^{\thinspace m}}

\section{Introduction}

Orthogonal polarization modes (OPMs) are the most striking phenomenon observed  
in the radio pulsar polarization data (eg.~Stinebring et al.~1984,
\nocite{scr84, mar2015} Mitra et al.
2015; Hankins \& Rankin 2010; van Straten \& Tiburzi 2017) but also in FRBs (Niu et al.~2024; Jiang et al.~2024).
\nocite{nwj2024, jxn2024, hr10, vst17} %%Qu \& Zhang 2023). 
They have always been attributed to the birefringent properties of
magnetized plasma (eg. Ruderman \& Sutherland 1975; Barnard \& Arons 1986; McKinnon
2003; Verdon \& Melrose 2008; Wang et
al.~2010; Hakobyan et al.~2017; Oswald et al.~2023),
\nocite{rs75, ba86, mck2003, vm2008, wlh10, hbp17, okj2023} 
where they appear as possible solutions of the
convolved system of the equation of motion and Maxwell equations for the
charge-plasma-field system. In averaged pulse profiles the modes are often observed
overlapping (Noutsos et al~2015; Johnston et al.~2008, Wang et al.~2023)
\nocite{nsk15, jkm2008, whx2023}
 and tend to have similar strength, which can also occur  
in single-pulse data.
% ... (refs). 
There is no understanding of what makes the modes often nearly equal 
nor why and where in the profile one mode should exceed the strength of another. 

In this paper we calculate polarization profiles for a special case of fan
beam geometry, with a localized emission source quickly passing next to the
observer's sightline. 
%[stream?]
It is shown that even in single-mode calculations,  
averaging effects lead to the appearance of two orthogonal polarizations of similar amounts. 
In radio pulsar magnetospheres, a similar passage
is traced by plasma that follows bent magnetic field lines and emits
relativistically beamed curvature radiation. % (Ruderman \& Sutherland 1975\nocite{rs75}). 
The angular scale and polarization structure of such a beamlet (or microbeam)
are known in the vacuum case (Jackson 1975\nocite{jac1975}) although they
are prone to reprocessing by coherency or propagational effects (interactions such as inverse Compton
scattering may possibly magnify the beamlet's scale, see Dyks 2023 with the polarization
discussed in sect.~4 therein). 
The model of this paper (Sect.~\ref{model}) has been inspired by this classical pencil beam, although
it is based on different assumptions about coherency of integrated signal.
After explaining the nonbirefringent OPMs in 
Sect.~\ref{resu}, we apply the model to azimuthally-extended emission
regions in Sect.~\ref{lateral}, then we discuss physical implications 
in Sect.~\ref{disc}.

\section{The model}
\label{model}

\begin{figure}
\begin{center}
\includegraphics[width=0.47\textwidth]{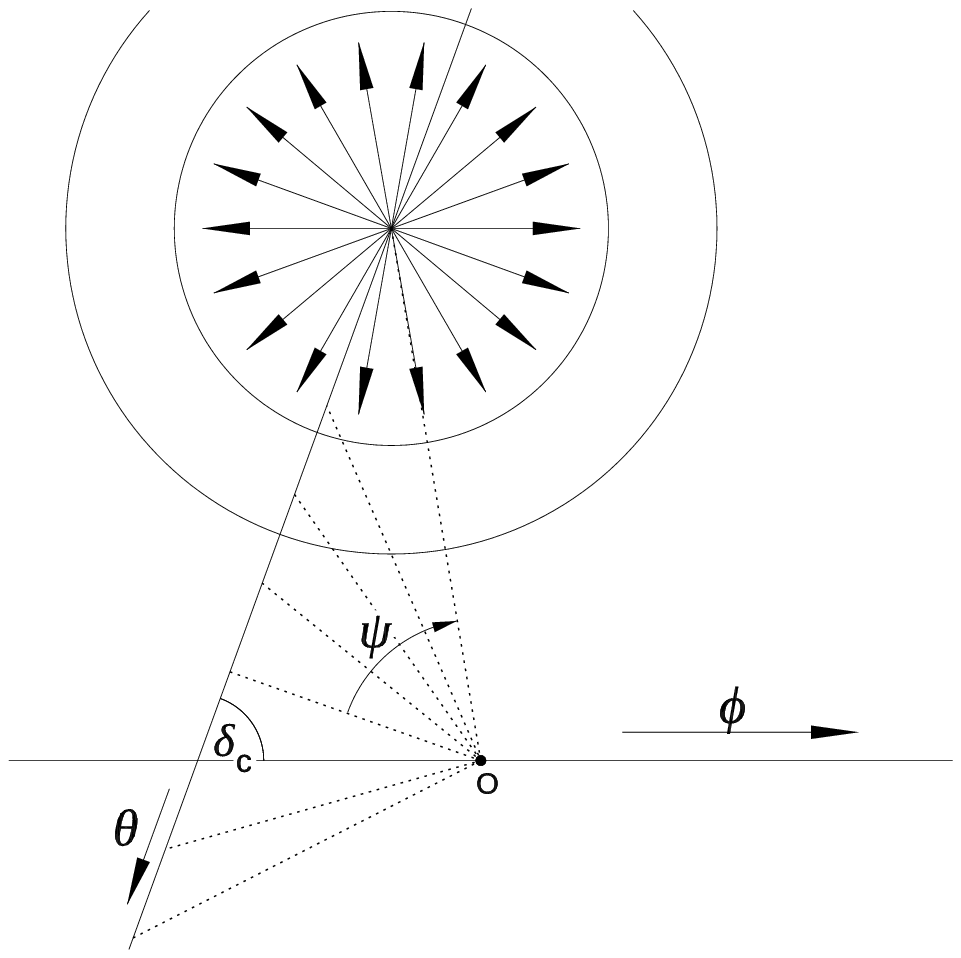}
\end{center}
\caption{Geometry of a beamlet passing in $\theta$ direction across the
horizontal path of sightline (currently at position `O'). The radial arrows
show the polarization direction within the entire beamlet. Neglecting the
slow sightline motion, the contributed polarization direction (at angle $\psi$)
is reversed 
as shown with dotted lines ($\psi$ follows an unresolved S-swing).}
\label{geom}
\end{figure}

We assume that a source of beamed radiation moves fast along the pulsar's
dipolar  $\vec B$-field line, its beam sweeping across the pulsar sky. 
Quasi-instantaneously, the radiated beamlet is assumed to have a narrow pencil-like shape
with a polarization direction that is radial around the beamlet's axis  
(the beamlet is shown face-on in Fig.~\ref{geom}). The
assumed radial structure of the polarization is natural for an axially symmetric
beamlet and more importantly it is consistent with the $(\vec k,\vec B)$ direction
of the ordinary polarization mode (the radial lines connect the tip of
the $\vec B$ vector at the center of the beamlet with the current observation
point O, i.e.~with the line of sight that is orthogonal to the page). 
\footnote{The polarization direction of the other (orthogonal) mode - that
would circle around the beamlet's axis - produces the same polarization angle curves 
(shifted by $90^\circ$).} 
The beamlet %(shown face-on in Fig.~\ref{geom}) 
is quickly moving in the
$\theta$ direction, ie.~away from the dipole axis. 
Because of the neutron star rotation, the observer's line of sight 
slowly moves horizontally in the direction of pulse longitude $\phi$.  
Assuming that the $\theta$ motion is relativistic, while the $\phi$ motion
is not, we neglect the latter, fixing the line of sight at a steady position 
corresponding to each pulse longitude. 

The beamlet is then swept past the sightline path at angle $\delta_c$, 
while contributing smoothly-changing polarization 
directions at each longitude $\phi$. The polarization
angle  $\psi$ is measured from a normal to the beamlet path, i.e. with
respect to the rotating vector model value (RVM value). 
The results are not very sensitive to the assumed intensity
profile of the beamlet.
% as long as its longitudinal and perpendicular dimensions are similar. 
Therefore, just for the sake of convenience,  
for the beamlet's emissivity we arbitrarily assume:
\begin{equation}
\eta = N(\theta, \phi)\eta_{\rm cr}
\label{beam}
\end{equation}  
where $\eta_{\rm cr}$ represents intensity of the curvature radiation (CR). 
To try various beam shapes we consider the CR emission either in the ordinary or
the extraordinary mode 
($\eta_{\rm cr}= \eta_{\rm cr,||}$ or $\eta_{\rm cr}= \eta_{\rm cr,\perp}$). 
The formulae for $\eta_{\rm cr}$ along with discussion of $||$ and $\perp$ index meaning 
are given in Dyks (2023, Sect.~4.5, eq.~7). \nocite{d2023} 
During the passage the intensity is rising up and dropping down, which is
taken into account through the radial (beamlet-centered) intensity profile $N$:
\begin{equation}
N = (1 + (\kappa/(1/\gamma))^2)^{-n}
\label{norfac}
\end{equation}
where $\kappa(\theta)$ is the instantaneous angular distance from the beamlet axis, 
and $\gamma$ and $n$ are parameters governing the radial emissivity scale 
(we take $\gamma=10$ for the beamlet to extend for a few degrees, and we usually assume
$n=6$; in the formulae for $\eta_{\rm cr}$ we fix the electron trajectory
radius at $\rho_c=10^5$ cm). 
This means that the observed signal
(collected at point O in Fig.~\ref{geom}) is dominated by the
contributions emitted when the beamlet axis is close to the line of sight.   
For simplicity it is assumed that longitude bins are wide enough to contain the entire
sweep, which lasts much shorter than the observational time sample. 
Because of this the sweeps that fall within two adjacent bins
are infrequent and are neglected. 

In our calculation we consider incoherent
superposition of radiative contributions at each longitude $\phi$. 
Thus, for each $\phi_j$ the beamlet is moved incrementally along $\theta_i$, with the
Stokes parameters calculated in the usual way: 
\begin{eqnarray}
I(\theta_i, \phi_j) & = &  \eta\\
Q(\theta_i, \phi_j) & = & \eta\cos(2\psi)\\
U(\theta_i, \phi_j) & = & \eta\sin(2\psi)
\label{stokes}
\end{eqnarray}
and summed over $\theta_i$ to $I(\phi)$, $Q(\phi)$ and $U(\phi)$:
\begin{equation}
I=\sum_{i}I(\theta_i, \phi_j), \ \ Q=\sum_{i}Q(\theta_i, \phi_j), \ \
U=\sum_{i}U(\theta_i, \phi_j)
\label{sums}
\end{equation}
The beamlet is assumed to be fully linearly polarized along the radial pattern
shown in Fig.~\ref{geom}, i.e.~in the $\vec k$-$\vec B$ plane.   

The physical basis for the model geometry cannot be directly attributed to
the CR microbeam on account of the coherent nature of the latter (see the
discussion in Sect.~\ref{disc}). The geometry can also be considered as
corresponding to a stream, because the temporal integration of the passage
is mathematically equivalent to
a spatial integration along a fan beam (also see Sect.~\ref{disc}).

\section{Nonbirefringent orthogonal modes}
\label{resu}

\begin{figure}
\begin{center}
\includegraphics[width=0.47\textwidth, height=0.35\textheight]{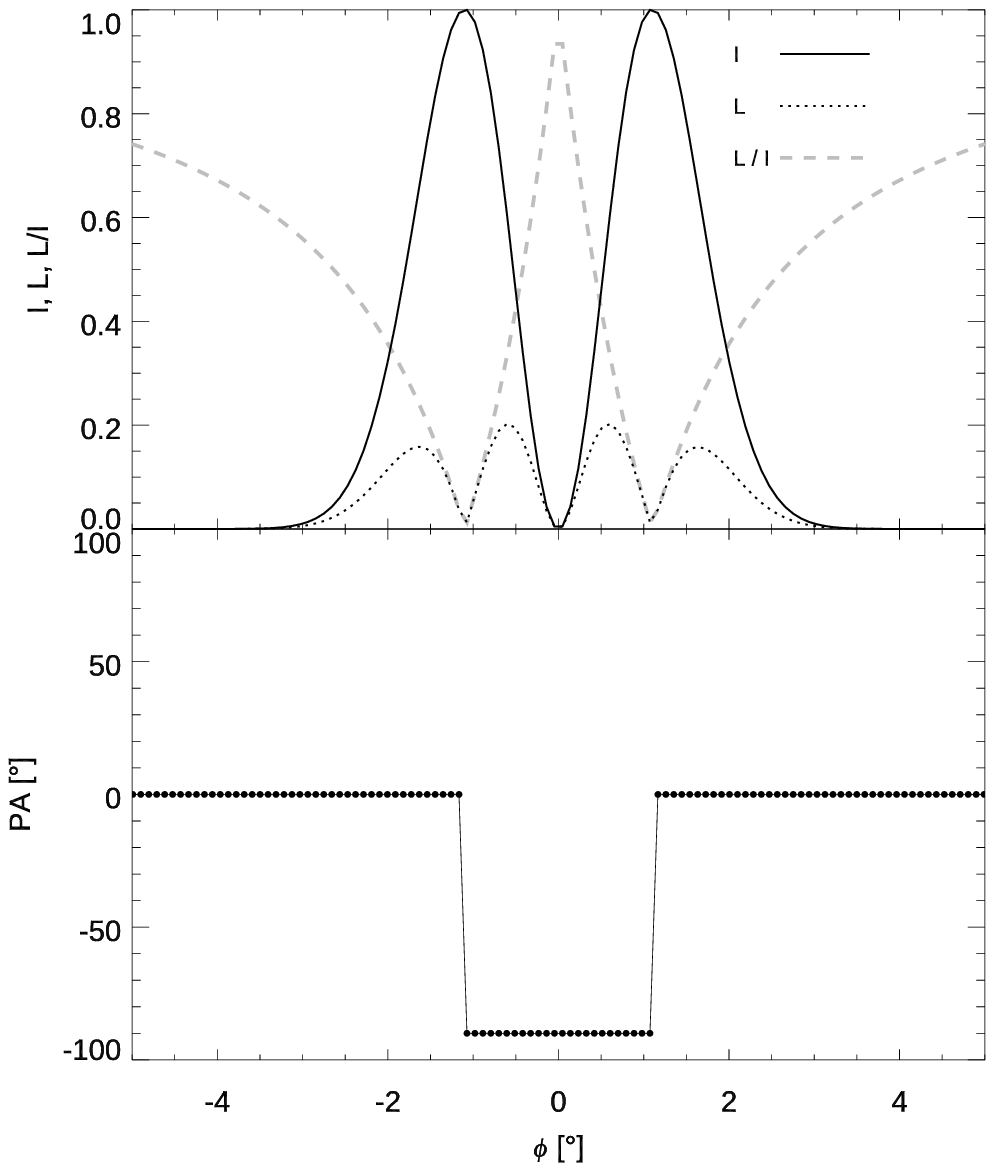}
\end{center}
\caption{Polarization profile for a beamlet with intensity 
of the O-mode CR microbeam ($\eta_{cr} = \eta_{cr,\parallel}$ in
eq.~\ref{beam}). 
Although a single polarization mode is used (polarized within the $\vec
k$-$\vec B$ plane) a clear OPM jump appears in the signal at pulse longitude
$\phi\sim \pm 1^\circ$. 
}
\label{jump}
\end{figure}

\begin{figure}
\begin{center}
\includegraphics[width=0.47\textwidth]{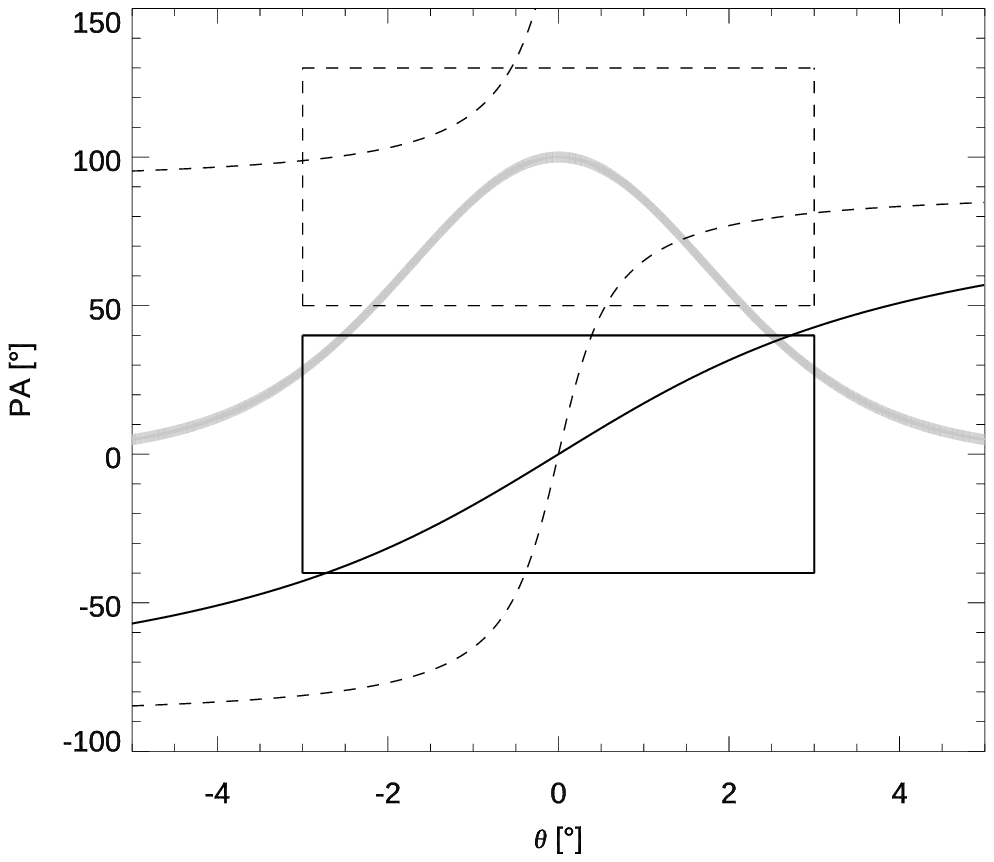}
\end{center}
\caption{Mechanism of the non-birefringent OPM jump. The intensity
(light grey) and PA (solid and dashed) are shown as a function of beamlet position
along its path. 
For the beamlet passing at a further distance from sightline, the PA follows the
flattish 
solid RVM-like curve that averages to the value of $\psi=0$. For a closer passage 
(steep dashed line) the peripheric values dominate and average to $\psi=90^\circ$,
as marked with the dashed rectangle.}
\label{cause}
\end{figure}

\begin{figure}
\begin{center}
\includegraphics[width=0.47\textwidth]{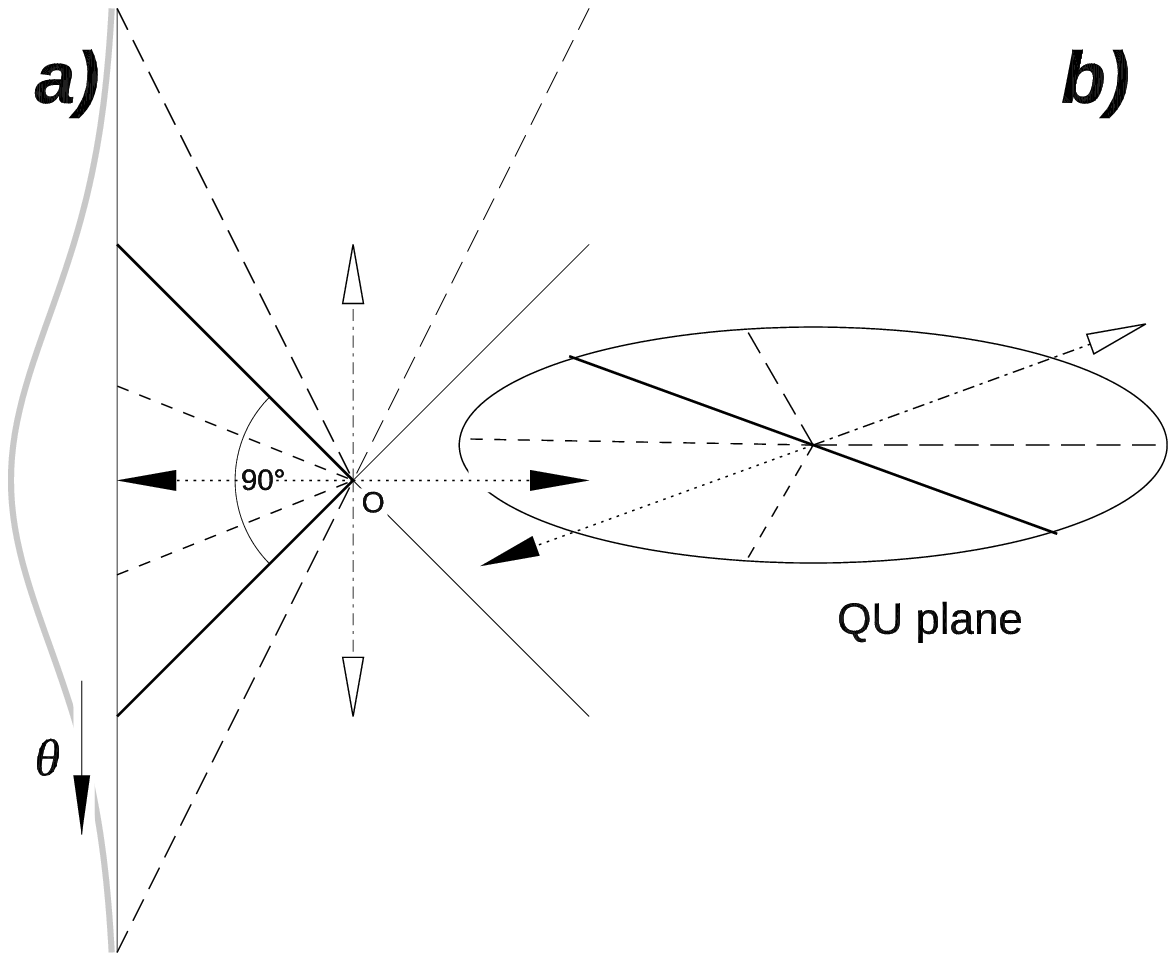}
\end{center}
\caption{Geometry of the non-birefringent OPM jump. 
Emitted intensity (grey line) is symmetrical with respect to the closest approach of the
beamlet to the line of sight at O. Each pair of symmetrical polarization directions 
within the solid line quadrant (eg. two short dashed lines) contributes
polarization parallel to the dotted arrows. 
Outside the quadrant (eg.~the pair of long dashed lines) the outcome is
orthogonal, ie. along the white-tip dot-dashed arrows. b) shows the
QU plane view.}
\label{show}
\end{figure}

Fig.~\ref{jump} shows the intensity $I$ (solid), linear polarization $L$
(dotted) and the
linearly polarized fraction $L/I$ (dashed) as well as the polarization angle (PA,
bottom) calculated for $\delta_c = 90^\circ$ 
%(Fig.~\ref{geom}) 
and $\eta_{\rm cr}= \eta_{\rm cr,||}$ (bifurcated
beamlet). This is a single-mode calculation with every contributing signal 
%contributing to the line of sight 
having the polarization along the $\vec k$-$\vec B$ plane (as
in Fig.~\ref{geom}). In spite of that, clear OPM jumps appear near  
$\phi=\pm1^\circ$.  

This can be understood by considering how the contributed PAs change while
the beamlet is moving in $\theta$, ie.~across the sightline path. 
During the passage, the contributed polarization direction essentially turns around,
hence the PA follows an RVM-like curve (solid and dashed curves in 
Fig.~\ref{cause}). For reference, the intensity profile is shown there in light-grey. 
The average PA measured in a single observed time sample is dominated by the
central (closest-approach) values, represented by the rectangles.    
 When the beamlet is passing by at a larger
distance from the sightline, the PA curve is rather flat (has a moderate slope, solid line case) 
and concentrated around $\psi=0$, therefore, the PA averages to $\psi=0$. 

However, when the line of sight approaches the crossing point of the beamlet trajectory 
(while the neutron star is rotating) the contributed PA curve is steeper
(dashed line). The central steep part of the ${\rm PA}(\theta)$ curve becomes so
narrow, that the average is dominated by the peripheric PA values (in the wings of the
grey intensity
bump). Since in the incoherent case the PA is periodic in $180^\circ$,  
the PA averages to $\psi=90^\circ$ as marked by the dashed rectangle. 

The origin of the OPM jump visible in Fig.~\ref{jump} is non-birefringent.  % Thus 
The reason for the jump is that the steep part of the PA$(\theta)$
curve does not weigh sufficiently in the average. 
As can be seen in Fig.~\ref{jump}, in the split beamlet case the radiation
is weakly polarized, with $L/I$ reaching $0.2-0.4$ only in the wings where
$I\sim0.2-0.4I_{\rm max}$ 
 
An important feature is that the contributed intensity is (quasi)symmetrical around
the point of closest approach to the line of sight. Therefore, it is useful
to consider polarization from pairs of symmetrical locations, as
shown in Fig.~\ref{show}a with pairs of same-style lines. Whenever such a pair is
contained within the solid-line quadrant facing the closest approach point
(as is the case for the short-dashed lines) the pair averages to the PA marked with the dotted
arrows 
(the lines in the pair fall symmetrically in one half of the QU plane as shown in Fig.~\ref{show}b).  
Beyond the indicated quadrant, another pair of lines (long dashed) falls
symmetrically within the opposite half of the QU plane, and contributes to the
orthogonal (antipodal) polarization (dot-dashed arrows with white tip). Which
mode is observed depends on which of these combined polarization states is brighter
(whether the white tip or black tip arrow on the QU plane is longer). 
To see this in the normal space, polarization directions must be
extended both sides with respect to O, as exemplified by the long-dashed line in
Fig.~\ref{show}a. In other words, the phenomenon of orthogonal modes in our model
results from the symmetry of the intensity in the beamlet passage. That
symmetry defines quadrants of orthogonality such as those centered at the O
point in Fig.~\ref{show}a. 
The strength of each mode (length of the black-tip versus white-tip arrow)
changes with the distance of the observation point from the passage (sweep) 
trajectory. The quadrant in Fig.~\ref{show}a subtends a section of different
size within the sweep trajectory, and the OPM jump occurs when the cumulated
intensities (within and beyond the quadrant) are equal.

\begin{figure}
\begin{center}
\includegraphics[width=0.47\textwidth, height=0.35\textheight]{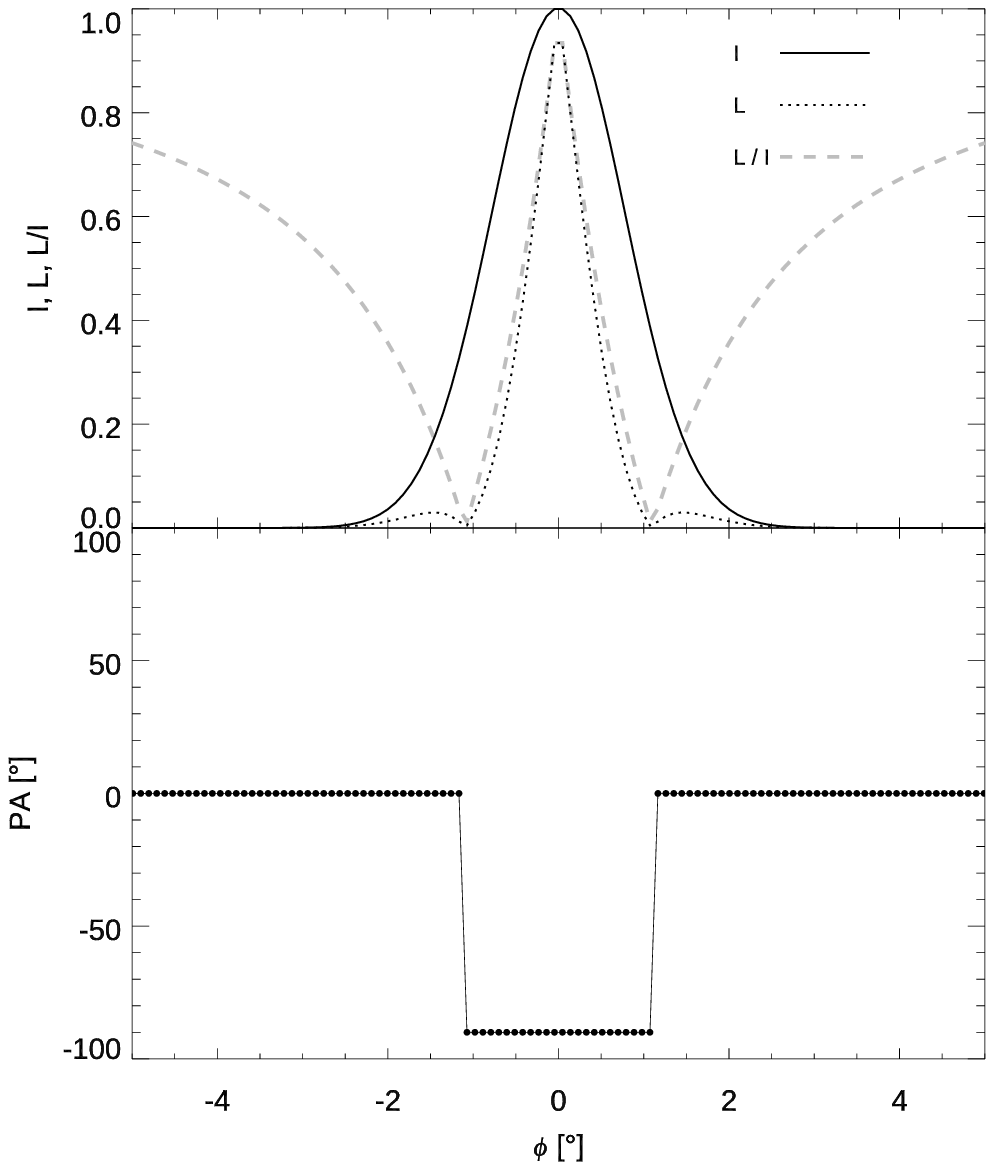}
\end{center}
\caption{Same as in Fig.~\ref{jump} but for a filled-in beamlet 
($\eta_{cr} = \eta_{cr,\perp}$ in eq.~\ref{beam}). The polarization
characteristics are similar except from the additional highly polarized part
in the center.
}
\label{pencil}
\end{figure}

Fig.~\ref{pencil} shows the case of $\eta_{\rm cr}= \eta_{\rm cr,\perp}$
with the beamlet in the form of the filled-in cone as shown in
Fig.~\ref{geom}.\footnote{%!!!
In the case of CR, the polarization direction in such a beamlet is
orthogonal to the one shown in Fig.~\ref{geom}, however, to allow for
arbitrary nature of the beamlet (see Sect.~\ref{disc} below) we maintain the radial
orientation also for the filled-in beam.} 
%and  ignore this $90^\circ$ shift on the vertical scale in PA plots.}
The polarization profile of the filled beam (Fig.~\ref{pencil}) is similar to
the previous case, except that now a highly polarized signal appears in the very center of
the $I$ profile. The PA curve exhibits a similar OPM jump since it is governed
by variations of the contributed PA and $I$ in the $\theta$ direction
(orthogonal to the horizontal axis of Fig.~\ref{pencil}).

Qualitatively, the results of Figs.~\ref{jump} and \ref{pencil} bear considerable resemblance to radio pulsar
polarization data. The model produces similar amounts of two orthogonal modes, ie.~two orthogonal 
PAs that used to be termed `modes' observationally. One of
the modes tends to dominate near the peak of intensity 
which sometimes occurs in the known data 
(e.g.~in PSR B2020$+$28 the weaker mode is flanking the second peak, 
fig.~16 in Mitra et al.~2025). 
\nocite{mar2015}
The filled beam case offers a wide range of polarized fraction, which
is indeed typically observed in single pulse data for many objects. 
The fast turning-around of polarization direction (dotted lines in
Fig.~\ref{geom}) resembles the fast PA rotations that have been resolved on a
slower timescale for PSR B1919$+$21 (Cao et al.~2025; Primak et al.~2022;
Manchester et al.~1975). 
\nocite{cjd2025, pts2022, mth1975}
However, the range of PA variations here is limited to $180^\circ$.  
Depending on the time scale and exact moment of the passage, the emission 
may extend for more than a single time sample. In such a case 
the PA can assume intermediate values, different than $0$ and $\pm 90^\circ$. 
Extended distributions of PA values are often observed around the average
RVM tracks in real single-pulse data (eg.~Young \& Rankin 2012).
\nocite{yr12} 
In the current purely linearly-polarized case the model is free from
circular polarization effects. To take them into account, at least a partially coherent 
gathering of beamlet's radiation would have to be considered during the passage. 

%Fig.~\ref{}

\section{Laterally extended emission regions}
\label{lateral}

\begin{figure*}
\begin{center}
\includegraphics[width=0.26\textwidth, height=0.23\textheight]{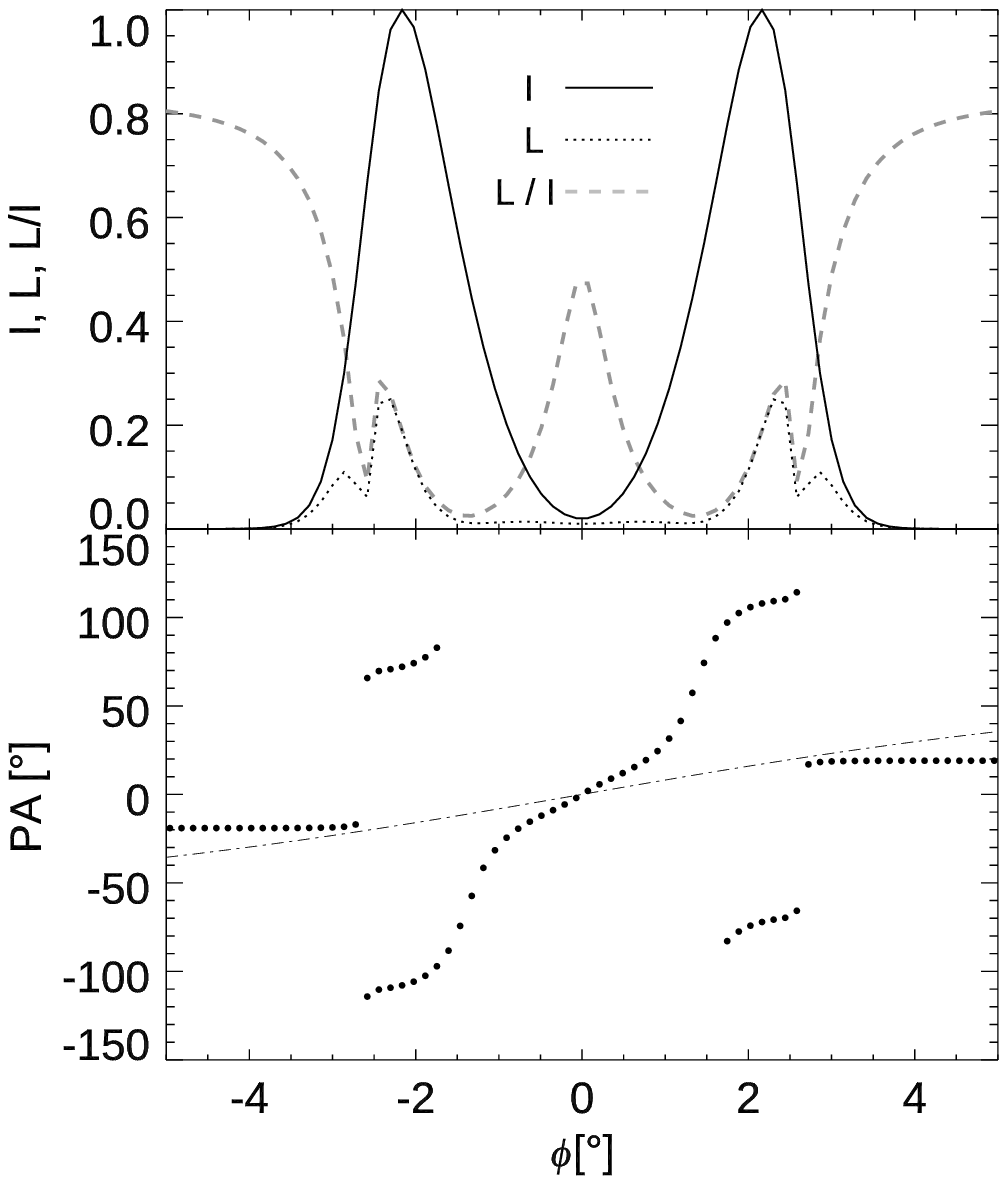}
\includegraphics[width=0.24\textwidth, height=0.23\textheight]{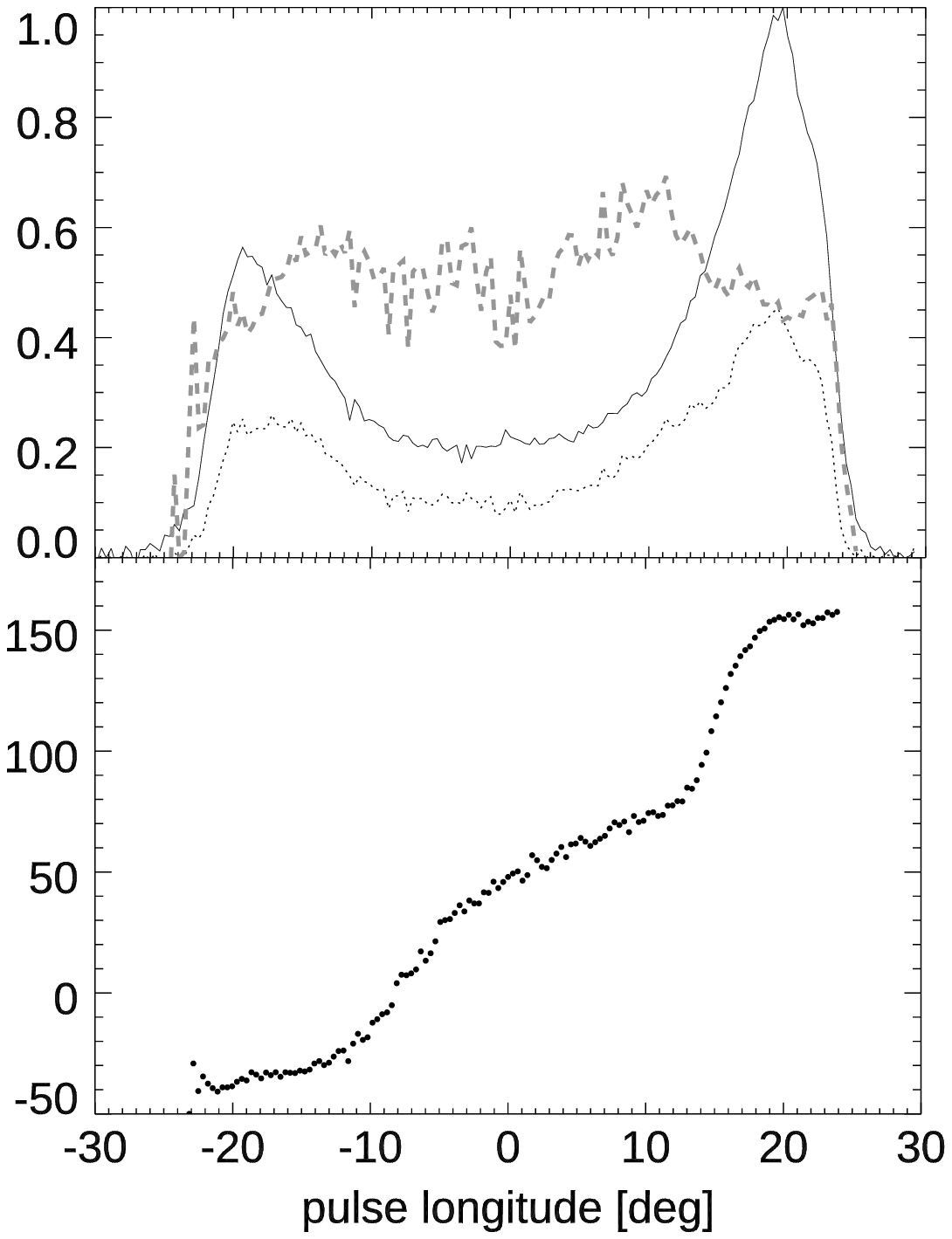}
\includegraphics[width=0.24\textwidth, height=0.23\textheight]{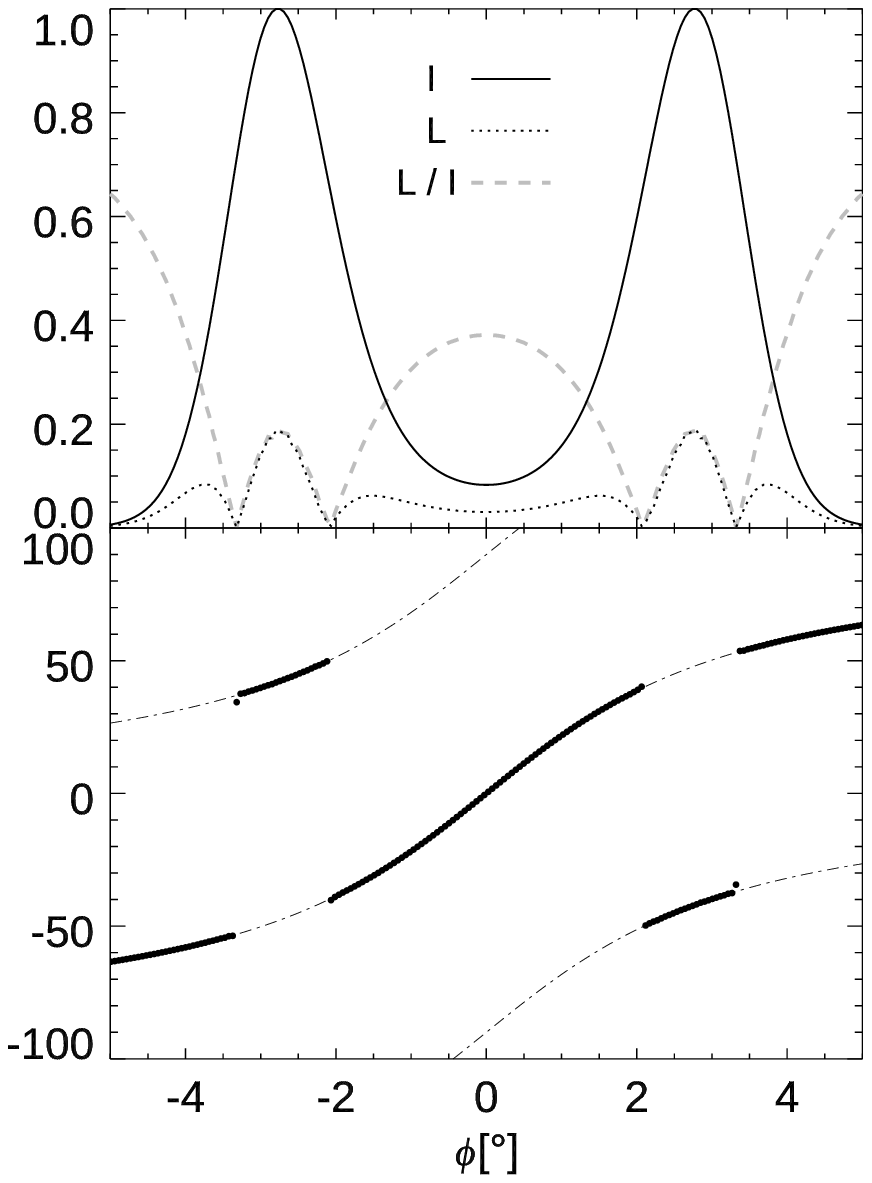}
\includegraphics[width=0.23\textwidth, height=0.23\textheight]{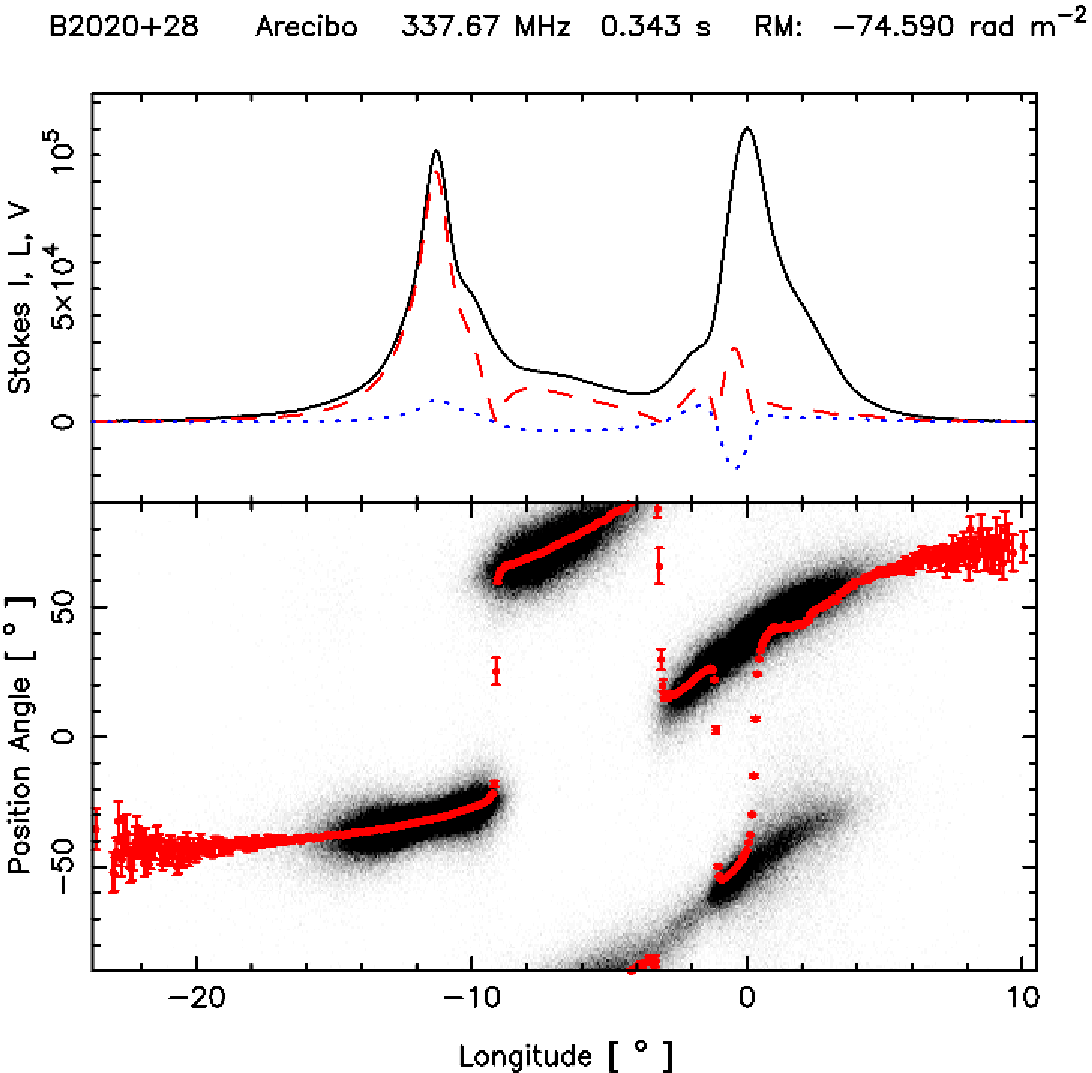}
\end{center}
\caption{Selected model results compared to data. 1st column: case of a wedge with bright
edges, producing a step-like PA curve (dot-dashed line shows RVM). 2nd column: Arecibo data on PSR
B1913+16 (Weisberg \& Taylor 2002). %\nocite{wt2002}
3rd
column: a hollow cone case characterized by narrow intervals of another mode under peaks and high
bridge polarization. 4th column: Arecibo data on PSR B2020+28 (Mitra et
al.~2015). %Thin dot-dashed line in model PA plots shows RVM. 
PA zero point is arbitrary. For more details see text. 
}
\label{comp}
\end{figure*}

Here we present examples of model results for two generic geometries of emission region:
a narrow wedge of non-negligible width (thus a stream or fan-beam) and the traditional hollow cone.
The model of Sect.~\ref{model} is extended threefold: 1) by adding some lateral extent $\Delta
\phi_m$ in magnetic azimuth $\phi_m$ (measured around the dipole axis), 2) by allowing for different brightness of the individual streams 
at a given $\phi_m$, and  3) including another longitudinal normalization factor $F(\theta)$, 
eg.~to form a narrow emission ring of the hollow cone.

Fig.~\ref{comp} (left) presents the case of a wedge with
$\Delta\phi_m=40^\circ$ for the 
case of the filled-in beamlet, $\beta=7^\circ$, and 200 streams evenly distributed within
$\phi_m\in(-20^\circ,20^\circ)$. The azimuthal normalization was changing
as $|\phi_m|^{2.5}$ (a wedge gradually brightenning towards the edges).
%linearly from 1 at $\phi_m=\pm2$ to 0.05 in the center. 
There was no longitudinal weighing ($F(\theta)=1$) and a narrow beamlet with
$\gamma=50$ was used. %and electron trajectory curvature $\rho_c=10^5$ cm was used.  
Such a wedge produces the type of PA behaviour observed for PSR B1913+16 (Weisberg \&
Taylor 2002), \nocite{wt02} i.e.~a step-like PA curve with a flat central part and orthogonal
mode in the periphery. The three mode zones -- with another mode flanking
the central mode - can arise in the model without the propagational effect of
the O-mode refraction. However, the agreement is qualitative only, with
details of $I$ and $L/I$ profiles not reproduced.
In general, the results are sensitive to the longitudinal
normalization that was absent here. In particular, the longitudinal variations
of emissivity make the OPM jumps gradual.

The third column of Fig.~\ref{comp} presents the traditional hollow cone
case calculated for a filled beamlet ($\gamma=50$) and 200 streams (magnetic azimuths).  
The cone had the angular radius of $7.5^\circ$ and a Lorentzian profile with
the thickness comparable to the beamlet width. 
This result shows interesting similarity
to the right-hand side of the Arecibo profile for PSR  B2020+28 (rightmost panel, Mitra et al.~2015). 
\nocite{mar2015} 
The polarized fraction
shows a small bump coincident with the trailing intensity peak, with the PA
visiting the orthogonal mode for a short while. The overall polarization
is low but increasing in the bridge between the components, which is often
observed in several double profiles (e.g.~Young \& Rankin
2012\nocite{yr12}).
%\footnote{\bf The direction of OPM jumps is different than in data, which is likely not important
%since it depends on which side of the V pole the intermediate polarization
%state is passing at the jump (an arbitrarily small miss of the V pole can
% make the azimuth on the QU plane either increasing or decreasing at the OPM jump, Dyks 2020).}
%\nocite{d2020}

The result can be understood by noting that an arc-shaped `bottom' part of the dipole-axis-centered ring
(the hollow cone) can be considered as a bent fan beam, albeit with
a horizontal orientation (thus with polarization similar to that of Fig.~5). 
The traverse of sightline through the cone then corresponds to moving into the stream,
passing its middle, then retreating back to the original side. While flux superposition makes
$L/I$ low near the trailing peak maximum, the high polarization in the
bridge corresponds to the peripheric high polarization of the stream
(Fig.~\ref{pencil}, $\phi\simeq\pm2^\circ$).

On the other hand, the result is inconsistent with the left side of the
B2020+28 profile, which is highly polarized. 
The geometry of the passage through the leading
component may be different, however, 
along with the above-discussed wedge case, this
probably means that the model does not fully grasp
all effects that shape $L/I$. 

While the introduced  model is based on the symmetry intrinsic to the stream or fan-beam
geometry, the third column of Fig.~\ref{comp} implies that
azimuthally-extended (conal) regions can also produce similar polarization
effects. Thus it is not the radial structure of diverging dipolar field lines, but elongation of
(any) emitter that leads to OPMs through the described symmetry effects. Indeed, the
radial (dipolar) field structure does not matter in Fig.~\ref{show} - it is
rather the symmetry of the emitter (of the $I(\theta)$ profile) with respect to the
line of sight (and the assumed polarization in the $(\vec k, \vec B)$ plane).
In general, a laterally extended emitter defines symmetry
that corresponds to a base PA value (e.g.~the RVM value) whereas
the question of which OPM track is traced depends on how much flux is integrated within 
pairs of quadrants of orthogonality in the vicinity of the line of sight 
(each pair aligned with either the filled or open arrows in Fig.~\ref{show}a).

\section{Discussion}
\label{disc}

The physical picture behind the geometry of Fig.~\ref{geom} is not fully understood.  
The standard equations for the curvature microbeam, eg. Jackson (1975),
\nocite{jac1975} 
involve the ultra-fast beamlet passage, however, we 
have mostly used them for illustrative purpose and for practical reasons, 
since they provide the known intensity envelope 
and can be easily scaled eg. through the value of Lorentz factor $\gamma$. 
However, the signal of CR microbeam is coherent and it is precisely the turn-around of
polarization direction shown in Fig.~\ref{geom} (dotted lines)  
that corresponds to the elliptically polarized CR pulse with the well-known antisymmetric V
profile. To collect CR contributions in the incoherent way, one would have
to assume that wave oscillation phase corresponding to different stages of
the passage is randomized somewhere on the way towards the
observer (which contradicts the coherent nature of the radiation, 
%and is anyway meaningless given that 
{\bf as} it is the entire passage that gives rise to the
elliptically polarized wave). 

The passage of the CR microbeam lasts for a time interval of the inverse of
observed frequency, ie.~$\Delta t \sim \nu_{\rm obs}^{-1}$ which is $1$ ns
at about $1$ GHz. Time samples in typical radio observations are much
larger (several tens of microseconds). Therefore, it is possible that the
beamlet of Fig.~\ref{geom} originates from a spatially much larger entity
observed on a longer timescale. This would allow for incoherent PA
contributions along the sightline, while still allowing for coherency at a
shorter time scale. The radial orientation of polarization could
either correspond to plasma density gradients (Benacek et al.~2025)
\nocite{bjp2025}
or could represent the usual single
polarization mode (eg. the $\vec k$-$\vec B$ plane). The observations of PSR
B1919$+$21 (Cao et al.~2025; see also Manchester et al.~1975) \nocite{cjd2025, mth1975}
definitely show that pulsar signal contains
rotations of polarization state that are both very fast and at the same time
 much longer than the microscopic time scale. 

Mathematically, the temporal integration of radiation from the beamlet passage is
equivalent to a spatial integration along a steady narrow stream.
The lack of
coherency between radiative contributions from different places along the
stream may seem to be more natural than for the localized emitter associated with the
beamlet. 
However, in the stream case it must be assumed that each length increment of the
stream emits the beamlet of Fig.~\ref{geom} to reach from the emission point
to the line of sight. Assuming that the strong magnetic field forbids
any considerable velocity spread within the stream (excluding effects such
as large pitch angles of gyration) it is difficult to justify the non-zero beamlet size otherwise than through
the microbeam of a radiative process.

Without clear microphysical picture it is difficult to 
claim that observed OPM jumps have the non-birefringent origin (or which of them do). 
The early applications of Sect.~\ref{lateral} look promising, though
they require more study. Nevertheless, the presented model bears some of the key geometric features of
the magnetospheric plasma motion and provides a rare example of geometry 
that produces clear OPM jumps resulting from just signal averaging. 
The model shows that nonbirefringent OPM jumps are possible, at least within
the kinematic (geometric) considerations. If the presumed
conditions are fulfilled in the pulsar magnetosphere 
then the averaging effects can mimic real OPMs and complicate the usual birefringent picture 
of pulsar polarization.  

\section*{Acknowledgments}
This research was funded by National Science Centre, Poland, grant 
no.~2023/49/B/ST9/01783.  
LS acknowledges support from the U.S. National Science Foundation and the Smithsonian Institution. 
JD thanks Dipanjan Mitra for permission to reproduce part of Fig.~19 from Mitra et
al.~(2015). We also thank Joel Weisberg for the average pulse data on B1913+16. 
For the purpose of Open Access, the author has applied a CC-BY public
copyright licence to any Author Accepted Manuscript version arising from
this submission.

\section*{Data Availability Statement} 
This paper refers to published data.

\bibliographystyle{mnras}
\bibliography{listofrefs2}

\end{document}